\documentclass[a4paper]{jpconf}
\usepackage{graphicx}
\usepackage{amssymb,amsmath}

\usepackage{citesort}

\begin{document}

\title{An instability of unitary quantum dynamics}

\author{Jasper van Wezel}

\address{Institute for Theoretical Physics, IoP, University of Amsterdam, 1090 GL Amsterdam, The Netherlands}

\ead{vanwezel@uva.nl}

\begin{abstract}
Instabilities of equilibrium quantum mechanics are common and well-understood. They are manifested for example in phase transitions, where a quantum system becomes so sensitive to perturbations that a symmetry can be spontaneously broken. Here, we consider the possibility that the time evolution governing quantum dynamics may be similarly subject to an instability, at which its unitarity spontaneously breaks down owing to an extreme sensitivity towards perturbations. We find that indeed such an instability exists, and we explore its immediate consequences. Interpretations of the results both in terms of extreme sensitivity to the influence of environmental degrees of freedom, and in terms of a possible fundamental violation of unitarity are discussed.
\end{abstract}

\section{Singular limits}
Instabilities in theories of physics are often signalled by the presence of a singular limit~\cite{Berry:2002}, which describes the extreme sensitivity of a physical situation to perturbations. The stability of a ball balanced on the top of a hill, the turbulent flow of a zero-viscosity fluid, and the chaotic motion of Saturn's moon Hyperion, are all examples of situations in which a singular limit heralds the fact that one physical concept has reached the bounds of its usefulness, and is surpassed by another~\cite{Berry:2002}. In many cases, the culprit is the thermodynamic limit, implementing Anderson's celebrated expression that `More is Different'~\cite{Anderson:1972}. 

\begin{figure}
\begin{center}
\includegraphics[width=0.75\columnwidth]{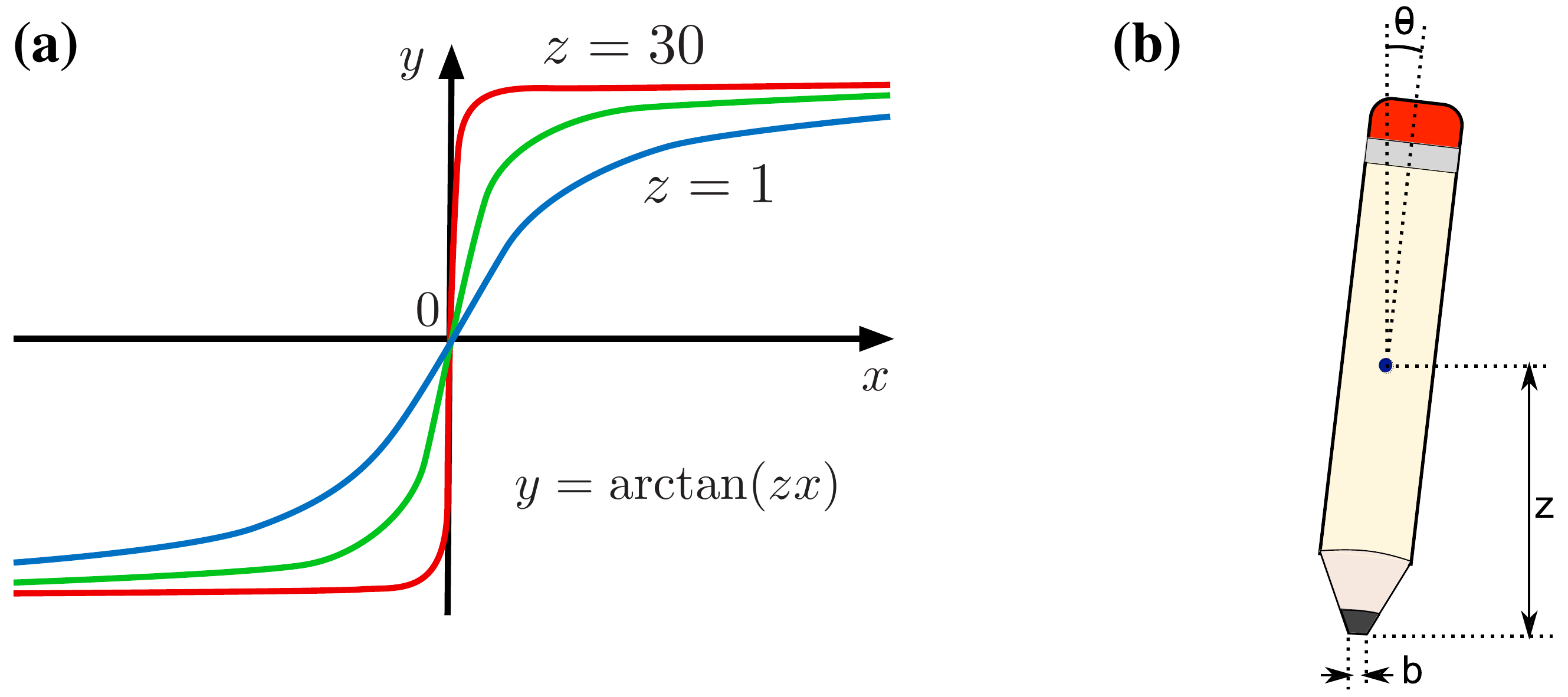}
\label{pencil}
\caption{{\bf (a)} The function $y=\arctan(z x)$ for different values of the parameter $z$. The value of the function at $x=0$ is infinitely sensitive to perturbations in the limit $z\to \infty$. {\bf (b)} A pencil balanced on its tip. The pencil is not perfectly sharp, having a blunt area with diameter $b$, and it is not perfectly balanced, dipping by an angle $\theta$. The fate of the center of mass height $z$ depends on the order in which the limits $b\to 0$ and $\theta \to 0$ are taken, signalling an extreme sensitivity to perturbations of $\theta$ in the limit of infinite sharpness.}
\end{center}
\end{figure}

In all of these cases, the singularity of the involved limit, and the associated sensitivity to perturbations can be expressed mathematically by the presence of a set of non-commuting limits. As an elementary example, consider the function $y=\arctan(z x)$, where $z$ is a parameter (see figure~\ref{pencil}). For any finite value of $z$ the function is perfectly smooth, and the limit of $y(x)$ as $x$ goes to zero is simply zero. However, upon taking the limit $z\to\infty$, the arctangent turns into a step function, and at that point the limit of $y(x)$ as $x$ approaches zero from above, is one. We thus find that the order in which the limits are taken matters:
\begin{align}
\lim_{z\to\infty} \lim_{x\downarrow 0} \arctan(z x) &= 0 \notag \\
\lim_{x\downarrow 0} \lim_{z\to\infty} \arctan(z x) &= 1.
\end{align}
The limit $z\to\infty$ is said to be singular, because it is qualitatively different from the situation with any finite $z$, no matter how large $z$ gets. At the same time, the value of $y(0)$ in the singular limit $z\to\infty$ is very sensitive to perturbations. In fact, the value of $y(\epsilon)$ differs qualitatively from $y(0)$, even for infinitesimally small $\epsilon$. Such an extreme sensitivity to even infinitesimally small perturbations is characteristic of singular limits throughout physics.

A physical example of this idea can be seen when we consider a sharp pencil being balanced on its tip (see figure~\ref{pencil}). The state of the pencil is encoded in the height $z$ of its center of mass. After sharpening the pencil such that it only has a blunt area of diameter $b$ left at its tip, and balancing it to within an angle $\theta$ of being perfectly upright, the fate of the pencil is determined by a singular limit:
\begin{align}
\lim_{b\to 0} \lim_{\theta \to 0} z &= 1 \\
\lim_{\theta \to 0} \lim_{b \to 0} z &= 0.
\end{align}
That is, although it may be possible to perfectly balance a somewhat blunt pencil, starting with a sufficiently sharp pencil ($b \to 0$) makes it infinitely sensitive to perturbations so that even an infinitesimal perturbation away from $\theta=0$ causes the pencil to tip over. In practice, it is therefore impossible to balance a sharp pencil on its tip. Notice that we can draw this conclusion about real pencils, even if neither of the limits is ever actually realised. The practical implication of the singular nature of the limit is that there is an extreme sensitivity to perturbations, and this sensitivity is manifested already for very (but not infinitely) sharp, well (but not infinitely-well) balanced pencils.

\section{Instabilities of equilibrium quantum mechanics}
The most common occurrence of singular limits in quantum mechanics, is in the spontaneous breakdown of symmetries. In this case, the thermodynamic limit presents a boundary to the applicability of the usual equilibrium description, and the state of the system is instead determined by infinitesimally weak perturbations. To see how this instability arises, it is instructive to consider the example of a harmonic crystal, described by the Hamiltonian:
\begin{align}
\hat{H} = \sum_j \frac{\hat{P}_j^2}{2 m} + \frac{1}{2} m \omega^2 \left( \hat{X}_j-\hat{X}_{j+1}\right)^2.
\label{crystal}
\end{align}
Here, $\hat{X}_j$ and $\hat{P}_j$ are the position and momentum of particle $j$, which has mass $m$ and is connected to neighbouring particles by a harmonic potential with natural frequency $\omega$. It is easy to check that this Hamiltonian commutes with the operator for total momentum $\hat{P}_{\text{tot}}=\sum_j \hat{P}_j$. In fact, this remains true even in the presence of anharmonic potentials, proportional to higher powers of the distance between neighbouring particles. The fact that the Hamiltonian for a crystal and the operator for total momentum commute, implies that they share a common set of eigenstates. The eigenstates of total momentum are all plane waves, with an even distribution of the centre of mass position over all of space. According to the usual rules of quantum mechanics, the state of the crystal in thermal equilibrium is given by a thermally weighted mixture of eigenstates of the Hamiltonian. It thus follows that for any temperature whatsoever, the equilibrium centre of mass position of a crystal is undefined, and its fluctuations diverge: $\langle \hat{X}_{\text{c.o.m.}}^2 \rangle = \infty$. In other words, under equilibrium conditions, any crystal ought to be spread out over all of space. 

Of course, this situation is a direct consequence of the presence of translational symmetry, and the reason that localised objects do exist in our everyday world is that translational symmetry can be spontaneously broken in the thermodynamic limit. In that limit, the ground state of the Hamiltonian becomes extremely sensitive to perturbations of the form:
 \begin{align}
\hat{H}' = \hat{H} + \frac{1}{2} N m \Omega^2 \left(\hat{X}_{\text{c.o.m.}} - X_0 \right)^2,
\label{SSB}
\end{align}
where $\Omega$ is the natural frequency of a harmonic potential  which tends to localize the centre of mass coordinate $\hat{X}_{\text{c.o.m.}} $ of the crystal with total mass $N m$ at the position $X_0$. The strength of the perturbation is given by the steepness of the potential, encoded in the natural frequency $\Omega$, which is considered to be infinitesimally small. In the limit of a crystal containing infinitely many particles, the ground state in the presence of any perturbation is qualitatively different from the ground state with $\Omega$ strictly zero. In that thermodynamic limit, even an infinitesimally small perturbation suffices to completely localize the centre of mass position of the crystal, and entirely suppress its fluctuations~\cite{JvW:AmJPhys07}:  
\begin{align}
\lim_{N\to \infty} \lim_{\Omega \to 0} \langle \hat{X}_{\text{c.o.m.}}^2 \rangle &= \infty \\
\lim_{\Omega \to 0} \lim_{N \to \infty} \langle \hat{X}_{\text{c.o.m.}}^2 \rangle &= 0.
\end{align}
As before, the harmonic crystal is only strictly localised in the presence of an infinitesimally small perturbation, if it is truly infinitely large. For a crystal of finite size, the non-commuting limits instead signal an extreme sensitivity to small, but non-zero, perturbations. Since the strength of the perturbation required to localise an object consisting of $N$ particles scales as $1/N$, objects as large as tables and chairs are susceptible to even the smallest perturbation in their environment, and in practice can never avoid its symmetry breaking influence.

Notice that any symmetric state can be written as a linear combination or wave packet of localised states. In the case of the harmonic crystal, the symmetric ground state of $\hat{H}$ can be formed from a superposition of localised states which are related to each other by rigid translations of the centre of mass position. These rigid translations are the symmetry operations corresponding to the translational symmetry of the Hamiltonian, and are generated by the total momentum operator. They do therefore not affect the internal degrees of freedom of the crystal, and form only a very small subset of all available states. Seen from the opposite point of view, one could equally well say that the localised ground state of the harmonic crystal in the presence of a perturbation, contains only a small subset of all eigenstates of the symmetric Hamiltonian. That is, the localised wave packet only contains states with differ in their values of the total momentum, but which have exactly equal quantum numbers for all internal degrees of freedom (phonon excitations). Moreover, precisely these total momentum eigenstates which make up the localised wave packet become degenerate in the thermodynamic limit, enabling the formation of a superposition with only an infinitesimal cost in energy, provided by an infinitesimally weak perturbation. The sparsity of the states contributing to this process as well as the collapsing nature of this tower of states in the thermodynamic limit, has earned them the designation of `thin spectrum' states~\cite{Anderson:1972,JvW:AmJPhys07,Anderson:1952,vanWezel:2005,vanWezel:2006}.

The extreme sensitivity to perturbations signalled by the presence of a singular limit, and the central role played by a collapsing tower of thin spectrum states are features which are present whenever a continuous global symmetry is spontaneously broken. A description precisely mirroring the above analysis can be given for the symmetry breaking observed in antiferromagnets, Bose-Einstein condensates, superconductors, and any other system with a spontaneously broken continuous symmetry~\cite{Anderson:1972,JvW:AmJPhys07,Anderson:1952,vanWezel:2005,vanWezel:2006,Kaiser:1989,Kaplan:1990,vanWezel:2008,vanWezel:2008ka,Birol:2007vw}.

\section{An instability of quantum dynamics}
The ubiquity of the singular thermodynamic limit in equilibrium quantum mechanics and the resulting extreme sensitivity of ground states to perturbations, suggests that the time evolution operator describing the dynamics of quantum objects in the thermodynamic limit may be similarly susceptible to infinitesimal disturbances. The defining symmetry of quantum dynamics is its unitary nature. The question we would thus like to pose is:
\begin{center}
``{\em Is the unitary time evolution implied by Schr\"odinger's equation} \\ stable {\em against infinitesimally small perturbations?'}' \\
~\\
Or equivalently: \\
~\\
``{\em Can unitarity be spontaneously broken?'}'
\end{center}
To answer this question, we follow the same procedure as that for the usual, static symmetries: an infinitesimally small perturbation is added to the Hamiltonian, and if this qualitatively affects the dynamics in the thermodynamic limit, the corresponding non-commuting set of limits signals the presence of an instability~\cite{vanWezel:2010fo,vanWezel:2011fy,vanWezel:2008prb}. 

For concreteness, consider a quantum Hamiltonian $\hat{H}$ and the time evolution generated by it: $i \hbar \partial / \partial t | \psi \rangle = \hat{H} |\psi\rangle$. When exploring infinitesimal additions to $\hat{H}$, it is clear that perturbations of the form $V \hat{O}$ with $V$ real and $\hat{O}$ a Hermitian operator will never break the unitarity of time evolution. To obtain non-unitary evolution, one needs to add an explicitly non-Hermitian perturbation. The simplest way to do this is to consider a perturbation of the form $i V \hat{O}$, with $V$ real and $\hat{O}$ Hermitian. Of course such perturbations are not allowed in the quantum mechanical description of any closed system. The study of spontaneous unitarity violations in closed quantum systems is therefore by definition a study of the stability of quantum dynamics towards perturbations {\em beyond} quantum theory~\cite{vanWezel:2010fo,vanWezel:2011fy,vanWezel:2008prb}. An alternative interpretation could be to treat the perturbed dynamics as describing the perfectly quantum mechanical, but effective dynamics of only part of a full system. It is well-known that the influence of the ignored environmental degrees of freedom can on average be modeled by a non-Hermitian term in the dynamics of the open quantum system~\cite{Caldeira:1983,Caldeira:1985,Unruh:1989}. We will comment on the implications of both of these interpretations after first considering the dynamics resulting from the presence of an infinitesimal unitarity breaking term.

As we are looking for a perturbation which will have an effect in the infinite particle limit $N \to \infty$, even for infinitesimally small strength of the perturbation $V\to 0$, it is clear that we need to consider perturbations which couple to an observable $\hat{O}$ that scales with the number of particles. Only in that way can the product of the two remain finite if the correct order of limits is taken. A natural candidate for such an extensive observable is the order parameter in a system which breaks a continuous symmetry. Choosing the perturbation $\hat{O}$ to be an order parameter moreover has the additional advantage that even the non-unitary time evolution will conserve energy in the thermodynamic limit. The reason for this, is that the order parameter only has non-zero matrix elements between states within the thin spectrum, which all collapse onto the ground state energy in the limit $N\to \infty$. In other words, a non-unitary term proportional to the order parameter field of a symmetry-broken system will affect the global value or orientation of the order parameter, but it cannot excite any of the internal degrees of freedom. 

For concreteness, consider again the harmonic crystal of equation~\eqref{crystal}. The unitarity breaking field in that case couples to the centre of mass position (the order parameter) of the crystal~\cite{vanWezel:2010fo,vanWezel:2011fy,vanWezel:2008prb}:
\begin{align}
i \hbar \frac{\partial}{\partial t} | \psi \rangle = \left[ \frac{\hat{P}_{\text{tot}}^2}{2 N m} + i \frac{1}{2} N m \Omega^2  \left( X_{\text{c.o.m.}}-X_0 \right)^2 \right] |\psi\rangle.
\label{non-unitary}
\end{align}
Notice that the unitarity breaking field in this expression is an imaginary version of the symmetry breaking field in equation~\eqref{SSB}, and that in the unitary part we focus on the collective motion of the crystal as a whole only, since the order parameter field will not affect the internal dynamics.

\begin{figure}
\begin{center}
\includegraphics[width=0.85\columnwidth]{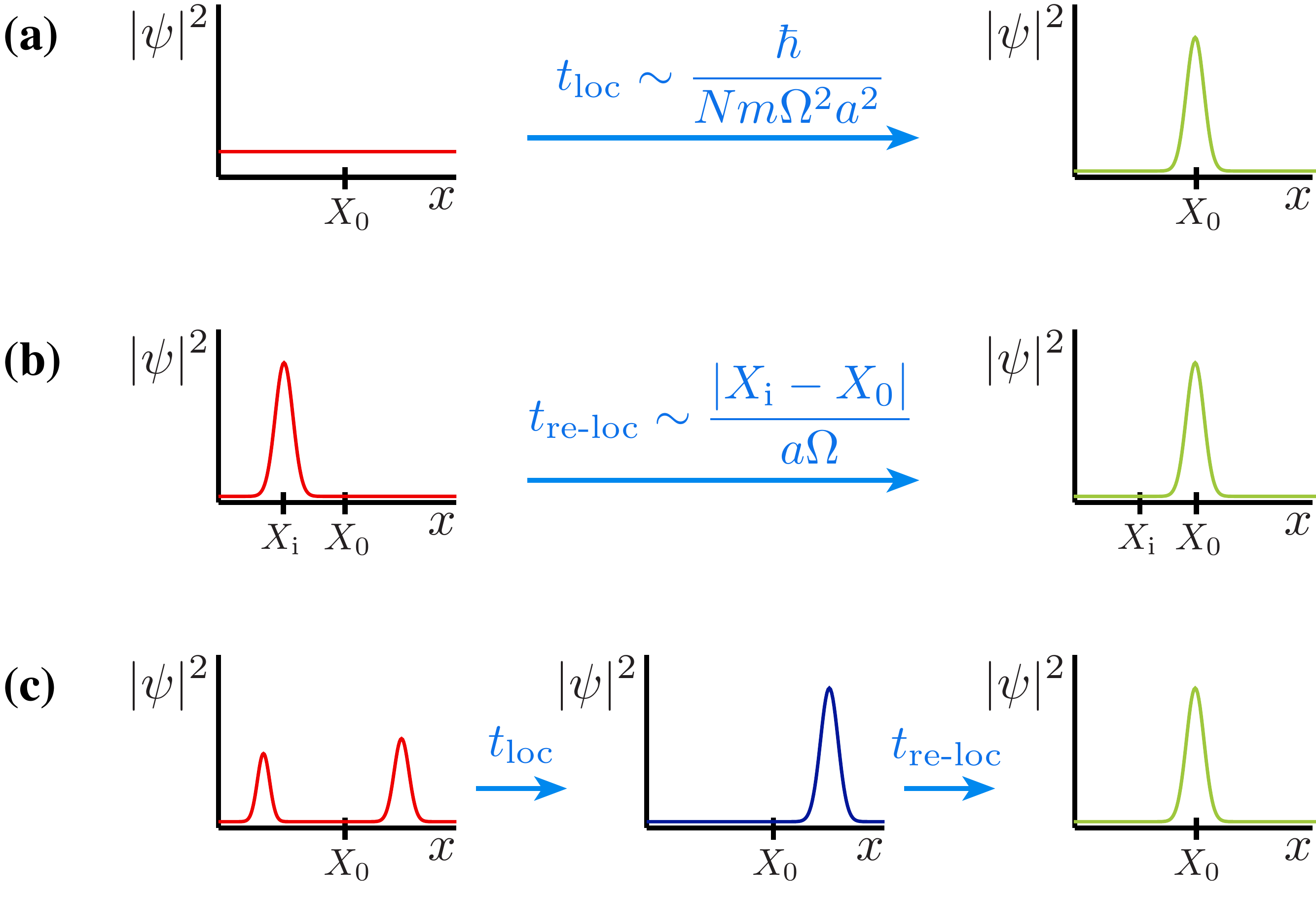}
\label{evo}
\caption{Sketch of the dynamics resulting from the non-unitary time evolution defined by equation~\eqref{non-unitary} for different initial conditions. {\bf (a)} Starting from a completely delocalised initial state, the wave function will be localised by the non-unitary perturbation within the time $t_{\text{loc}} \sim \hbar / ( N m \Omega^2 a^2)$. {\bf (b)} Starting from an already localised initial state at position $x=X_{\text{i}}$, the wavefunction will be re-located to position $x=X_0$ within the time $t_{\text{re-loc}} \sim |X_{\text{i}} - X_0 | / (a \Omega )$. {\bf (c)} A superposed initial state combines the two types of dynamics, and will first be reduced to a localised wave function at a single position within time $t_{\text{loc}}$, after which it will re-locate to position $x=X_0$ within the time $t_{\text{re-loc}}$.}
\end{center}
\end{figure}

The effect of the non-unitary dynamics defined by equation~\eqref{non-unitary} depends not only on the time evolution operator, but also on the initial state it is applied to. To characterise the possible effects, consider the three different initial conditions displayed on the left side of figure~\ref{evo}: (a) a uniformly spread, zero-momentum wave packet with equal amplitude everywhere in space, (b) a sharply localised wave packet at a position different from $x=X_0$, and (c) a superposition of two sharply localised wave packets with different amplitudes and different distances from $x=X_0$. The first two initial states can be seen as the extreme cases representing the ground states of the symmetric Hamiltonian and a symmetry broken, localised crystal respectively. The third initial state is an interpolation between these extremes. The initial states can be straightforwardly numerically integrated forward in time~\cite{vanWezel:2010fo}, resulting in the final states at late times displayed on the right of figure~\ref{evo}. Notice that during the time evolution defined by equation~\eqref{non-unitary}, the norm of the total wave function is not conserved. This is not a problem, since in quantum mechanics only relative amplitudes lead to observable predictions. All the usual interpretations of the laws of quantum theory remain consistent and well-defined, as long as expectation values are taken relative to the instantaneous norm of the wave function, so that:
\begin{align}
\langle \hat{O} \rangle (t) \equiv \frac{ \langle \psi(t) | \hat{O} | \psi(t) \rangle }{ \langle \psi(t) | \psi(t) \rangle}.
\end{align}

The eventual fate of the wave function at infinite time $t\to \infty$ is found to be a localised state at $x=X_{0}$ regardless of the initial state. This can be easily understood, as the non-Hermitian term in the evolution of equation~\eqref{non-unitary} suppresses any component of the wave function away from $x=X_0$. For the completely delocalised initial state, which is an eigenstate of the total momentum, the suppression of weight by the non-unitary term is the dominant process, and the time scale on which the wave packet is localised can be estimated by considering the evolution in the absence of any Hermitian term. A straightforward calculation then shows that the spread of an initially delocalised wave packet will be reduced to a single lattice spacing, $a$, within the time $t_{\text{loc}} = \hbar / ( 2 N m \Omega^2 a^2)$. Direct numerical integration shows that a non-zero kinetic energy term will affect the prefactor of this localisation time, but not its proportionality to $\hbar / (N m \Omega^2 a^2)$~\cite{vanWezel:2010fo}. Notice that this proportionality immediately implies that the thermodynamic limit and the limit of vanishing perturbation do not commute:
\begin{align}
\lim_{N\to \infty} \lim_{\Omega \to 0} t_{\text{loc}} &= \infty \\
\lim_{\Omega \to 0} \lim_{N \to \infty} t_{\text{loc}} &= 0.
\end{align}
We thus find that infinitely large quantum systems are sensitive to even infinitesimally small non-unitary perturbations, which will instantaneously localise any delocalised state. In other words, the unitary quantum dynamics of a delocalised state is {\em unstable} against non-unitary perturbations. For finite sized objects this means that the dynamics of large delocalised quantum systems will be dominated entirely by the influence of arbitrarily small non-unitary terms, if any exist.

Starting from an already localised initial state, the dynamics resulting from equation~\eqref{non-unitary} behaves differently. In this case, the initial wave packet has zero amplitude at position $x=X_0$, and the non-unitary term can therefore not amplify the relative weight of that component. The initial time evolution is instead dominated by the Hermitian kinetic energy, which tends to spread a localised state. Disregarding the non-unitary term entirely, it is straightforward to see that the time it takes for an initially localised wave packet to be spread to a width $|X_{\text{i}}-X_0|$, is given by $t_{\text{spread}} = 2 N m |X_{\text{i}}-X_0|^2 / \hbar$. Once the wave function has been spread over this distance, it develops a non-zero amplitude at the position $x=X_0$, and the localising effect of the non-unitary term takes over once again. The combined time scale needed for the wave packet to first spread in space and then localise again, can be estimated by the geometric mean of the individual contributions. This gives a re-location time for the initially localised state of $t_{\text{re-loc}} = \sqrt{t_{\text{spread}} t_{\text{loc}}} = |X_{\text{i}}-X_0| / (a \Omega)$. Once again, the direct numerical evaluation of the dynamics confirms that the proportionality of the re-location time to $|X_{\text{i}}-X_0| / (a \Omega)$ is maintained also if both terms in the evolution operator are simultaneously present ~\cite{vanWezel:2010fo}. Notice that in the case of the re-location time, the thermodynamic limit does commute with the limit of vanishing perturbation:
\begin{align}
\lim_{N\to \infty} \lim_{\Omega \to 0} t_{\text{re-loc}} &= \infty \\
\lim_{\Omega \to 0} \lim_{N \to \infty} t_{\text{re-loc}} &= \infty.
\end{align}
Even in the presence of a non-unitary perturbation, the dynamics of a large, localised quantum object is therefore {\em stable}. The non-unitary perturbation will not have an effect on a localised macroscopic state within any measurable time.

Finally, consider an initial state superposed with different weights over two positions, at different distances to $x=X_0$. In this case, the dynamics consists of both a localisation and a re-location. At first, the component closest to $x=X_0$ is amplified by the non-unitary term relative to the further component. As before, this results in a localised wave packet, but this time the centre of mass is determined by the position of the amplified component of the initial wave function. The time scale over which the localisation takes place is again given by $t_{\text{loc}}$. Once the wave function is localised, the slower re-location process takes over, and the wave packet is re-located to $x=X_0$ within the time $t_{\text{re-loc}}$. Taking the thermodynamic limit and the limit of vanishing perturbation, we then find that the dynamics of an initially superposed state is {\em unstable} agains non-unitary perturbations. Even an infinitesimally small non-unitary perturbation will cause a spatially superposed macroscopic quantum state to collapse onto just one of its components, which then is stable agains any further effects of non-unitary terms in the dynamics.

\section{Interpretation and speculation}
So far, all presented results are exact. It is a mathematical fact that the time evolution generated by the quantum mechanical Hamiltonian is unstable against even infinitesimally small non-unitary perturbations in the thermodynamic limit. To see if this mathematical observation also has any physical consequences, we need to consider whether an arbitrarily small, but non-zero, unitarity breaking perturbation could exist even in principle. If we consider closed systems only, and if we additionally assume quantum mechanics to hold universally, then the answer is clearly that such perturbations cannot exist, and the mathematical instability described above is irrelevant to physics. Relaxing either one of the constraints however, non-unitary perturbations are no longer strictly forbidden, and we can speculate about possible origins of non-unitary dynamics.

If we consider open quantum systems for example, it is well known that the influence of an unobserved environment on the system of interest can in many cases be effectively described by a unitarity violating term in its dynamics~\cite{Caldeira:1983,Caldeira:1985,Unruh:1989}. If the environment tends to localise a macroscopic object (which is typically the case), the non-unitary evolution will be similar to that of equation~\eqref{non-unitary}. One should keep in mind however that the environment is generically unknown, unpredictable, and evolving uncontrollably in time. The position $X_0$ at which it tends to localise the object of interest is thus an unknown parameter, whose value moreover changes unpredictably with time. This effective behaviour can be modelled by taking $X_0(t)$ to be a randomly fluctuating variable, and considering the time evolution generated by:
\begin{align}
i \hbar \frac{\partial}{\partial t} | \psi \rangle = \left[ \frac{\hat{P}_{\text{tot}}^2}{2 N m} + i N \gamma  \left( X_{\text{c.o.m.}}-X_0(t) \right)^2 \right] |\psi\rangle.
\label{decoherence}
\end{align}
Here $\gamma$ depends on the coupling of the system to the environmental degrees of freedom, and the scaling of the interaction with $N$ has been made explicit. Assuming the environment to be infinitely large, the randomly fluctuating variable $X_0(t)$ may be modeled by white noise over the entire range of available positions. In this interpretation of the origin of non-unitary evolution, the implication of the singular thermodynamic limit is that it is impossible even in principle to isolate macroscopic objects from their environments. Large objects are sensitive to infinitesimally small interactions with the environment, which will instantaneously affect their dynamics.

An alternative way of arriving at a similar description of time evolution for macroscopic quantum objects, is to consider closed systems only, but allow for the possibility that quantum theory may not be an exact description of nature. It has been proposed for example, that the diffeomorphism invariance underlying general relativity is incompatible with the unitary nature of quantum dynamics, and that this must result in either one or both of these principles being violated at the borderline between gravity and quantum physics~\cite{Penrose:1996}. In one particular proposal, dimensional analysis is used to argue that this borderline may lie close to the length scales at which we cease to be able tot observe quantum behaviour in experiments~\cite{Penrose:1996,Diosi:1987}. The non-unitary evolution discussed here, could then be interpreted as arising from a non-unitary modification of the laws of quantum mechanics due to gravity~\cite{vanWezel:2008jr,vanWezel:2011fy,vanWezel:2010fo}. The strength of the non-unitary perturbation in that case should be similar to the cross-over energy scale identified as the border between the two theories. Moreover, the principle of diffeomorphism invariance can be argued to cause any measure of a distance between objects in different components of a superposed wave function (in the position basis) to be inherently ill-defined~\cite{Penrose:1996}. As a poor man's approach, one could model such an ill-defined distance by adding a randomly fluctuating correction to the definition of the position operator: $\hat{X} \to \hat{X}-X_0(t)$. The fluctuations $X_0(t)$ do not correspond to any physical quantum field, but instead represent the fact that when the effects of diffeomorphism invariance start to play a role, distances between different components of the wave function cannot be predicted with certainty anymore~\cite{vanWezel:2011fy,vanWezel:2010fo}. The dynamics arising from such considerations of a macroscopic object in the region where quantum mechanics just begins to be affected by the principles underlying the theory of gravity, can then be written as~\cite{vanWezel:2011fy,vanWezel:2010fo}:
\begin{align}
i \hbar \frac{\partial}{\partial t} | \psi \rangle = \left[ \frac{\hat{P}_{\text{tot}}^2}{2 N m} + i \frac{1}{2} N m G \rho \left( X_{\text{c.o.m.}}-X_0(t) \right)^2 \right] |\psi\rangle.
\label{gravity}
\end{align}
Here $G$ is Newton's gravitational constant, $\rho$ is the density of the object, and $N m$ its total mass. $X_0(t)$ is again a random variable that can be modeled by white noise over the entire range of positions. If the non-unitary term is fundamental in the sense that it comes from physics outside of quantum mechanics and applies to the evolution of single-shot experiments (as opposed to the ensemble averaged description implicit in the case of an open quantum system), then the implication of the singular thermodynamic limit must be that the dynamics of macroscopic quantum objects is unstable by itself. Large enough objects can thus not be forced to evolve quantum mechanically under any circumstances, because they always probe the influence of physics beyond Schr\"odinger's equation.

The overall effect of the perturbed evolution in both equation~\eqref{decoherence} and~\eqref{gravity} is the same as that of equation~\eqref{non-unitary} which we considered before. That is, the unitary dynamics of an initially localised object is stable agains perturbations, while the dynamics of delocalised or superposed initial states are unstable. This time however, the infinite sensitivity to non-unitary perturbations does not lead to the deterministic outcome of a localised state at a predictable position. The reason is that the wave function component whose amplitude is effectively amplified by the non-unitary term in the dynamics, is the component closest to $x=X_0(t)$, which now randomly fluctuates in time. Starting for example from a superposed wave function over two initial locations, it may be that first one component is closer to $X_0(t)$, and hence is amplified, while an instant later the other component is selected, and so on. In this way, a randomly switching sequence of amplifications is applied to the two components, which terminates only when one of the components is suppressed to zero amplitude. The dynamics thus becomes a realisation of the ``Gambler's Ruin'' game of probability theory~\cite{Pearle:1976,Zurek:2005}. Using the principle of `envariance', it is possible to show that the only possible average outcome of a series of such evolutions, is the emergence of Born's rule~\cite{Zurek:2005,vanWezel:2010fo,vanWezel:2008prb}. That is, in each individual realisation of the dynamics of equation~\eqref{decoherence} or~\eqref{gravity} applied to the initial state $|\psi(0)\rangle = \alpha | x=X_{\text{1}} \rangle + \beta |x=X_{\text{2}} \rangle$, the outcome is unpredictable, but will be either $| x=X_{\text{1}} \rangle$ or $| x=X_{\text{2}} \rangle$. Repeating the same evolution many times, the proportion of times that the state $| x=X_{\text{1}} \rangle$ is realised will be given by $|\alpha|^2$, and the proportion of outcomes at $| x=X_{\text{2}} \rangle$ will be $|\beta|^2$. This conclusion is confirmed by direct numerical integration of equations~\eqref{decoherence} and~\eqref{gravity}~\cite{vanWezel:2011fy,vanWezel:2010fo}.

We thus find that spatial superpositions of macroscopic objects spontaneously collapse into the position basis according to the prescription of Born's rule. This collapse process is a direct consequence of the instability of their unitary quantum dynamics to even infinitesimally small non-unitary perturbations. The non-unitary term can be provided either by a modification to quantum mechanics originating from gravity, or from an effective description of the environment in an open quantum system.

\section{Experimental consequences}
The final result of the sensitivity of macroscopic objects to non-unitary perturbations is that such objects cannot retain a state of spatial superposition. The absence of macroscopic superpositions in the world around us is indeed easily verified. If an interaction between a microscopic quantum particle in a spatially superposed state and a localised macroscopic object is instantaneously turned on, this may instantaneously cause the macroscopic object to become entangled with the microscopic particle, and to enter into a superposed state. The subsequent very fast collapse to just one of the components, in accordance with Born's rule can then be interpreted as a measurement of the particle's position by the macroscopic object. Indeed, this process satisfies all of the usual requirements for a description of wavefunction collapse during a quantum measurement. To determine whether an evolution like the one imposed by equation~\eqref{decoherence} or~\eqref{gravity} is responsible for these observations however, requires the experimental observation of the actual dynamical evolution, rather than only its eventual outcome.

To observe the dynamics imposed by a non-unitary perturbation within experimentally accessible time scales, the object to be studied should not be too macroscopic, so that the effect of the perturbation is not too strong and hence $t_{\text{loc}}$ is large enough to be observable. On the other hand, it should also not be too microscopic, as otherwise the perturbation has no effect at all, and $t_{\text{loc}}$ becomes too large to be observed. The region in between, where collapse times can be expected to be of observable sizes, falls precisely in the region that is hard to access experimentally, and which is only very recently starting to be explored by modern experiments involving macroscopic superpositions~\cite{Penrose:1996,vanWezel:2011fy,Marshall:2003}.

If such experiments succeed in observing collapse dynamics, the scaling of the time scales involved may give some hint as to the origin of the non-unitary perturbation that underlies it. A more pertinent question however, will be whether such experiments can also distinguish between perturbations coming from the effective action of an unobserved environment, and perturbation coming from a fundamental modification of Schr\"odinger's equation. The expected dynamics resulting from these sources is practically indistinguishable, as shown by the similarity of equations~\eqref{decoherence} and~\eqref{gravity}. There is however an essential difference between the effective dynamics of an open system, and the possibly fundamentally non-unitary dynamics of a closed system~\cite{vanWezel:2011fy}: in the former case, the effective dynamics has to emerge from the neglect (tracing out) of some environmental degrees of freedom which are beyond the control of the experimental setup. The environment may be external or internal and involve spin, charge, massive, massless, or any other kinds of degrees of freedom, but it has to become entangled with the observed properties of the macroscopic object for the dynamics to become effectively non-unitary. Since the experiment by definition has no control over the environmental degrees of freedom, the entanglement between observed properties and the state of the environment cannot be undone once it has been created, and the collapse dynamics cannot be turned off once it has started. If on the other hand the non-unitary term is fundamental, no entanglement with environmental degrees of freedom is required. In that case, one can create a macroscopic superposition, let it begin its collapse dynamics, and then bring back both superposed components onto the initial position before the collapse is completed. Since there is no entanglement with the environment, the situation is just as it was before creating the macroscopic superposition, and this procedure can be repeated an arbitrary number of times. When entanglement with environmental degrees of freedom is present however, bringing the components of the superposed object back to their initial positions will not undo the entanglement, and hence a subsequent run of the experiment will be influenced by the entanglement left behind from previous runs. This will enhance the effectively non-unitary dynamics, which thus occurs on a shorter and shorter time scale as the experiment is repeated again and again. In this way, sequences of collapse experiments can be used to distinguish between effectively non-unitary dynamics in open systems, and truly non-unitary dynamics resulting from a fundamental modification of quantum theory~\cite{vanWezel:2011fy}.

\section{Conclusions}
We have found that the unitarity of quantum mechanical time evolution can be spontaneously broken, in precise analogy with the usual breaking of continuous symmetries under equilibrium conditions. The loss of unitary dynamics arises from an extreme sensitivity in the evolution of large quantum objects to even infinitesimally small non-unitary perturbations. If such perturbations can (effectively) exist, then regardless of their origin, it immediately follows that truly macroscopic objects can never be observed to evolve unitarily, as even the weakest unitarity violating interaction will instantaneously dominate its dynamics. The effect of this instability is that delocalised or spatially superposed macroscopic objects are instantaneously reduced to a single location. Objects which are already localised on the other hand, are insensitive to the non-unitarity, and will remain localised at the same position.  

Two possible sources of unitarity violating perturbations to quantum dynamics are the influence of an unseen environment, or a fundamental modification of Schr\"odinger's equation. In the former case, the dynamics of an open system entangled with environmental degrees of freedom can be effectively described by a non-unitary evolution after tracing over the environmental degrees of freedom. The effective dynamics generated in this way generically tends to localise large objects. For the latter case, it has been suggested that the conflict between general relativity and quantum physics originates in the fundamentally non-unitary nature of the diffeomorphism invariant theory of gravity. It may then be expected that the laws of quantum mechanics break down at an energy scale where gravity starts to appear, and that the first effect of this breakdown can be modelled by a non-unitary modification of quantum dynamics. In either case, the scaling of the non-unitary term with system size and the randomly fluctuating nature of the location at which it tends to localise large objects, is the same. The result of a non-unitary perturbation which includes random fluctuations, is the emergence of probabilistic dynamics for large objects. That is, the instability of unitary quantum dynamics in this case implies that a superposition of a macroscopic object over multiple spatial locations is instantaneously reduced to just one of these positions. The selection of which place the wave function collapses to, is a probabilistic process. The probability for a particular component of the initial state to be selected according to the dynamics generated by an infinitesimal, fluctuating, non-unitary perturbation, is given by Bron's rule.

Experimentally capturing the predicted non-unitary dynamics of macroscopic objects requires studying the dynamics of objects which are large enough to be influenced by the non-unitarity, but small enough not the instantaneously reduced to a localised state. Such mesoscopic quantum superpostions have very recently begun to become accessible in state of the art experimental setups. As further technological development allows these experiments to proceed to spatial superposition of large enough objects for the instability to non-unitary perturbations to become noticeable, it will be necessary to develop protocols which can distinguish between the effectively non-unitary, but strictly quantum mechanical, evolution induced by environmental degrees of freedom, and fundamental modifications of Schr\"odinger's equation. The presence of entanglement with environmental degrees of freedom in the former case, but not the latter, is an unavoidable difference that may be employed to tell them apart.

Regardless of the source of non-unitary contributions to the (effective) dynamics, the fact that spatial superpositions of sufficiently large quantum objects are inherently unstable against arbitrarily small perturbations, will affect experiments accessing quantum states at these scales within the near future.

\ack
This work was supported by a VIDI grant financed by the Netherlands Organisation for Scientific Research (NWO).

\section*{References}
\bibliography{DICE}

\end{document}